# Sub-nanosecond signal propagation in anisotropy engineered nanomagnetic logic chains

Zheng Gu[1,†], Mark E. Nowakowski[1,†], David B. Carlton[2], Ralph Storz[3], Mi-Young Im[4,5], Jeongmin Hong[1], Weilun Chao[4], Brian Lambson[6], Patrick Bennett[1], Mohmmad T. Alam[7], Matthew A. Marcus[8], Andrew Doran[8], Anthony Young[8], Andreas Scholl[8], Peter Fischer[4,9], and Jeffrey Bokor[1,*]

[1]*Department of Electrical Engineering and Computer Sciences, University of California, Berkeley, California 94720, USA*
[2]*Intel Corp., 2200 Mission College Blvd., Santa Clara, California 95054, USA*
[3]*Thorlabs Inc., 56 Sparta Ave., Newton, New Jersey 07860, USA*
[4]*Center for X-ray Optics, Lawrence Berkeley National Laboratory, Berkeley, California 94720, USA*
[5]*Daegu Gyeongbuk Institute of Science and Technology, Daegu 711-873, Korea*
[6]*iRunway, 2906 Stender Way, Santa Clara, California 95054, USA*
[7]*Intel Corp., 5200 NE Elam Young Pkwy, Hillsboro, OR 97124, USA*
[8]*Advanced Light Source, Lawrence Berkeley National Laboratory, Berkeley, California 94720, USA*
[9]*Department of Physics, University of California, Santa Cruz, California 94056, USA*

† Denotes equal contributions
* Corresponding author

**Abstract**

Energy efficient nanomagnetic logic (NML) computing architectures propagate and process binary information by relying on dipolar field coupling to reorient closely-spaced nanoscale magnets. Signal propagation in nanomagnet chains of various sizes, shapes, and magnetic orientations has been previously characterized by static magnetic imaging experiments with low-speed adiabatic operation; however the mechanisms which determine the final state and their reproducibility over millions of cycles in high-speed operation (sub-ns time scale) have yet to be experimentally investigated. Monitoring NML operation at its ultimate intrinsic speed reveals features undetectable by conventional static imaging including individual nanomagnetic switching events and systematic error nucleation during signal propagation. Here, we present a new study of NML operation in a high speed regime at fast repetition rates. We perform direct imaging of digital signal propagation in permalloy nanomagnet chains with varying degrees of shape-engineered biaxial anisotropy using full-field magnetic soft x-ray transmission microscopy after applying single nanosecond magnetic field pulses. Further, we use time-resolved magnetic photo-emission electron microscopy to evaluate the sub-nanosecond dipolar coupling signal propagation dynamics in optimized chains with 100 ps time resolution as they are cycled with nanosecond field pulses at a rate of 3 MHz. An intrinsic switching time of 100 ps per magnet is observed. These experiments, and accompanying macro-spin and micromagnetic simulations, reveal the underlying physics of NML architectures repetitively operated on nanosecond timescales and identify relevant engineering parameters to optimize performance and reliability.

**Main text**

Nanomagnetic logic (NML) is a computational architecture that promises ultralow energy dissipation per operation[1,2,3,4]. In NML, the magnetization of single-domain ferromagnetic thin-film islands are coupled by dipolar fields generated from adjacent islands. Single islands are typically ellipse or rectangular-shaped; this confines the magnetic easy axis to the major axis. When islands are arranged in a line parallel to the minor axis (magnetic hard axis), the nearest-neighbor magnetic dipolar field coupling imparts a preference for these neighboring islands to align anti-parallel, forming a spatial logical inverter[5]. An extended series of these closely spaced nanomagnets, called a chain, propagates binary information from one end to the other sequentially through a series of inversions, performing a function similar to conventional integrated circuit (IC) interconnects but with potentially lower dissipation energy per switching event[2,6]. These chains are a fundamental building block of NML architectures. To perform a logic operation, the magnetization state of an "output" magnet is determined by a majority vote of the magnetic state of three "input" magnets which surround the "output"; this majority logic gate has been experimentally demonstrated[2], and is projected to switch at energies near 1 eV (0.16 aJ), making NML a candidate for computing with energy dissipation approaching the fundamental thermodynamic limit[7].

To perform multiple, successive logic operations, the entire NML architecture (chains and majority logic gates) must be re-initialized after each operation. This resetting process is known as clocking and in this work we use pulsed nanosecond on-chip magnetic fields[8] to drive the magnetization of all nanomagnets in a chain to saturation along their magnetic hard axes. This places each magnet in an energetically unstable (null) state which, upon removal of the clock field, becomes coupled to a nearest neighbor magnet by the dipolar fields. The time-dependent relaxation from the null state in rectangular or elliptically shaped magnets, however, can also be affected by factors such as thermal fluctuations[9,10,11,12,13] and non-ideal magnetic anisotropies[14]. These aberrations can drive the magnetization of individual islands to spontaneously switch out of sequence forming an error that spoils the sequentially directed dipolar coupling which correctly propagates the input information. Error rates have been predicted to increase as a function of shorter clocking pulse lengths approaching nanosecond timescales in chains of 5 or more ellipse magnets[9] due to the non-deterministic settling of magnets far from the input. This has motivated efforts to prevent error nucleation in longer chains[10,11,12,13].

One method to increase the stability of the null state is to artificially engineer a metastable potential well along the hard axis by introducing a biaxial anisotropy component to each individual nanomagnet[11]. Two distinct methods can be used to incorporate this biaxial anisotropy: choosing a material with an intrinsic magnetocrystalline biaxial anisotropy[10,11] or fabricating lithographically-defined notches on both ends of each nanomagnet major axis[15,16,17,18] (Figure 1a, inset). In this work we choose the latter since it provides an opportunity to controllably tune the relative strength of the biaxial anisotropy. This method has been previously employed and was shown to reduce the influence of thermal fluctuations and random lithographic variations to an extent that error-free signal propagation along a chain of 8 nanomagnets was observed, driven by a quasi-static, adiabatic clocking process[13]. For NML to be a viable alternative logic architecture, fast, error-free operations, with speeds limited solely by the intrinsic magnetic relaxation time (order 100s of picoseconds), must be repeatable and reliable over successive clocking cycles[19,20,21,22]. Even though reliable high-speed operation has been predicted[11,12,13], it has yet to be experimentally studied in any NML architecture.

In this work, we study the critical elements (i.e. reliability and speed) required to produce successful repeatable signal propagation in chains with a fast cycling protocol. First, using magnetic full-field transmission soft x-ray microscopy (MTXM)[23] at the Advanced Light Source (ALS) synchrotron at Lawrence Berkeley National Lab, we statistically analyze the signal propagation reliability of nanomagnet chains, clocked by single nanosecond pulses, with varying magnet dimensions to identify regions of optimal signal propagation and verify that fast clocking can be reliably employed. Next, we directly image signal propagation dynamics in a chain of anisotropy-engineered nanomagnets with 100 ps time resolution using magnetic time-resolved photo-emission electron microscopy (TR-PEEM)[24,25,26] by clocking in sync with pulsed x-rays at a repetition rate of 3 MHz. Finally, we compare both experimental results with computational simulations. With a micromagnetic simulator, we examine the effectiveness and clocking behavior of nanosecond magnetic field pulses. Additionally, we use macro-spin simulations to examine relevant experimental parameters which affect the reliability and speed of signal propagation; these include temperature, nanomagnet dimensions, and the dipolar coupling strength. These simulations both validate our experimental results and suggest approaches for further technological improvement.

To study propagation reliability using MTXM, notched nanomagnet chains (Figure 1d) of varying length L, 150 nm wide and separated by 30 nm, are fabricated along an Al wire on x-ray transparent 100 nm thin $Si_3N_4$ membranes by a combination of e-beam and optical lithography, evaporation, lift-off, and wet etching techniques (See Methods). During the measurement, manually-triggered 3 ns clocking pulses generating an 84 mT on-chip field reset the chains (Figure 1a, See Methods). Dipolar signal propagation is initialized at the chain input[27] by magnets with shape anisotropy (indicated by the red oval in Figure 1d) designed to spontaneously and consistently orient along one direction of the easy axis after clocking (See supplementary section S1). For each length L, two nominally identical chains are fabricated with oppositely oriented input magnets (Figure 1e,f). Additionally, an ellipse-shaped 'block' magnet (indicated by the orange oval in Figure 1d) terminates each chain to stabilize the final nanomagnet after successful signal propagation[8] (See supplementary section S1). Magnetic contrast images after each pulse are generated by detecting x-ray magnetic circular dichroism (XMCD)[28] of the transmitted x-rays (Figure 1e, see Methods). The values of biaxial anisotropy ($K_b$) and uniaxial anisotropy ($K_u$) in Figure 2a are calculated using an analytical model and Object Oriented Micromagnetic Framework (OOMMF)[29] by simulating the field required to reorient magnets with lithographically defined anisotropies (See supplementary section S2). With the notch dimensions fixed ($K_b$ approximately constant), $K_u$ varies approximately linearly with the nanomagnet easy axis length (L) from 300-360 nm (Figure 2a) defining a variable ratio $K_b:K_u$ which parameterizes a lithographically-defined energy well for the metastable state (Figure 2a, inset) that the dipolar coupling fields must overcome to correctly reorient initialized magnets. We assess both the clocking stability and signal propagation distance as a function of this ratio.

Figure 2b plots the experimentally determined signal propagation distance of nanomagnet chains, when clocked as described above, as a function of L for two identically processed samples. The propagation distance is defined as the number of nanomagnets (starting at the input) over which the signal propagates correctly before encountering the first error. After propagation is complete, magnets which remain along their hard axes and magnets aligned parallel to either neighbor are both considered errors. Due to the static nature of this measurement, out-of-order switching, though technically errors, cannot be distinguished. For every value of L, four chains were each pulsed ten times. We observe a systematic propagation distance peak in sample 2 as $K_u$ is tuned by varying L. Error bars represent random variations in signal propagation on different trials. Sample 1 does not clearly show this peak, but rather contains a pair of chains with the same L showing perfect signal propagation every trial. Though sample-to-sample process variations lead to widely varying levels of performance, we demonstrate that significant improvements in signal propagation distance are possible with more consistent processing.

For nanomagnets with smaller L, the energy well set by $K_b:K_u$ is too deep for the dipolar coupling fields to overcome. After the clocking field is removed, the nanomagnets remain oriented along their hard axes, stabilized by the biaxial anisotropy (i.e. trapped in the metastable state). Figure 2c depicts the schematic magnetic configuration which is verified by an MTXM image of a chain fully "locked" in this manner (Figure 2d). For the larger values of L, $K_b:K_u$ is smaller. This reduces the dipolar field coupling required to reorient the magnets; however this also reduces the stability provided by the metastable state and makes it more susceptible to thermal excitations or anisotropies introduced by processing irregularities. Because of this, we observe reduced signal propagation distances in chains with longer nanomagnets. The high signal propagation regions in Figure 2b represent regions in each sample where $K_b:K_u$ is optimally tuned. Accordingly, Figure 2e depicts the schematic magnetic configuration and Figure 2f shows an MTXM image of a chain with perfect signal propagation.

Micromagnetic simulations with OOMMF provide further insight into the magnetic signal propagation observed in this experiment. In Figure 2g, we present simulated signal propagation lengths for chains of 12 magnets identical to the experimental dimensions and spacing (bounded by inputs[27] and blocks[8]; see supplemental section S1). Initially, we study magnet chains as they evolve from an ideal metastable state (ideal clock). Without thermal fluctuations (0K, black line) we observe a relatively large, sharply defined signal propagation region, spanning a variation of L of over 120 nm, of nearly perfect signal propagation. The final magnet in the chain is too strongly coupled to the block and remains oriented in the metastable state. This is an artifact of the simulation and hence a maximum signal propagation of 11 magnets is observed. This block coupling also creates signal propagation noise on the right side of the plot by acting as a nucleation site for errors. In actual NML architectures, we do not anticipate this behavior. At 300 K (red circles) stochastic thermal effects reduce the high propagation region to approximately 80 nm wide in L. This is because as $K_b:K_u$ is reduced, thermal energy can assist in prematurely reorienting magnets, creating errors.

Finally, we repeat the same simulations but no longer assume an idealized initialization condition. Instead, the simulated chains are subjected to a trapezoidal-shaped 3 ns clocked pulse (200 ps rise time, 300 ps fall time) approximating our experimental clock pulse (blue triangles). As with ideal clocking, signal propagation behavior is similar to idealized conditions and is a function of $K_b:K_u$, however the short clocking pulse does not provide the same degree of stability when switching to the metastable state. Hence, we find a smaller signal propagation region spanning only 50 nm in L. We also observe that the pulsed field, in conjunction with stochastic variability, produces a slightly non-uniform initialized state among individual magnets in a chain. This initialization randomness exacerbates the trend of reduced signal propagation at larger values of L, already present in ideally initialized chains. Nevertheless, despite fabrication-related imperfections, we observe a high signal propagation region spanning 20 nm in L in one sample (Figure 2b). The magnitude of this span is the same order of magnitude as with the simulated chains and suggests reasonable agreement between simulations and experiment. This both experimentally confirms previous work[11], predicting the existence of an optimized region for signal propagation in anisotropy-engineered nanomagnet chains, and also allows us to conclude that nanosecond current-generated field pulsing is an effective clocking mechanism that can be employed in an ultrafast measurement to reset chains over millions of cycles. Additionally, we predict (with micromagnetic simulations similar to Figure 2g) that error-free signal propagation using nanosecond clocking pulses is extendable to even longer magnet chains (24 magnets) that incorporate biaxial anisotropy (See supplementary section S3). This overcomes the error limitations exposed by exclusively using uniaxial magnets[9].

    To observe signal propagation in an anisotropy-engineered chain at nanosecond timescales, we use TR-PEEM. Notched nanomagnet chains with identical dimensions to the previous experiment (with lengths, L, from 300-500 nm) are fabricated on a Au wire on top of a Si wafer (Figure 1b) and mounted to a customized circuit board designed to apply nanosecond current pulses (Figure 1c, See Methods). The 2 ns current pulses (which generate 100 mT on-chip fields) are synchronized with 70 ps x-ray pulses at a repetition rate of 3 MHz. The time delay between the current and x-ray pulses is varied with a pulse delay generator. XMCD images (Figure 1f) are generated by aligning and dividing PEEM images illuminated with left and right circularly polarized x-rays. For a given delay, each image is an average of over 300 million clock cycles.

Figure 3a plots integrated photo-emission yield from a PEEM image of the wire versus time delay. The Lorentz force generated by the clock field deflects emitted photoelectrons during the clock pulse, shifting and reducing the intensity of the image. This confirms the generation of 2 ns magnetic field pulses and identifies the zero time delay point at the peak of the pulse. Figure 3b shows a series of time-resolved XMCD-PEEM images of a chain from the optimal region taken at delay times between 1 and 4 ns. Surface roughness and other lithographical irregularities may contribute to errors in this system[8,30,31,32] while random signal propagation errors from individual clocking cycles are averaged into each image. This leads to contrast "graying" from individual nanomagnets that do not reproducibly reorient in the same direction each cycle. Figure 3c illustrates our interpretation of the signal propagation in this chain. Imaging at zero delay confirms all nanomagnets are aligned along their hard axes (See supplemental section S4). One nanosecond after the clocking pulse peak the input magnet sets the initial condition for signal propagation; the remaining magnets remain aligned along their hard axis (perpendicular to the magnetic contrast direction). At 1.2 and 1.4 ns repeatable switching of the first few nanomagnets in the chain occurs. Between 1.4 and 1.8 ns we observe the sequential reorientation of nanomagnets in this chain. This confirms that signal propagation proceeds at a rate of approximately 100 ps per switching event, as predicted previously through computational NML studies[2,10,11]. In the 4 ns image the signal propagation is complete and appears error-free, however the dynamics measurement reveals an error nucleating out of sequence in the 1.8 ns image. The time-resolved technique we have used introduces the capability of interrogating the performance of individual nanomagnets during signal propagation, a feature not present in existing quasi-static imaging measurements. Experimentally evaluating the performance of NML chains complements existing time-resolved micromagnetic simulations (like the one demonstrated in supplementary section S3), offers realistic assessments of nanomagnet designs, and identifies systematic error nucleation sites and other architectural weaknesses resulting from environmental aberrations and lithographic limitations.

Additionally, this experimental observation verifies the fundamental mechanism upon which NML architectures are based. The high-speed regime of switching we have observed is governed by Landau-Lifshitz-Gilbert dynamics[10] and is distinct from the typically slower thermally-assisted switching that follows a modified Arrhenius model[19]. Successful signal propagation in this chain can be directly compared to previous work in which similar signal propagation was observed but driven by slow, adiabatic thermally-assisted switching[13]. This demonstrates that both 'fast' and 'slow' clocking of NML chains may be used, depending on the preferred system architecture for a given application.

To obtain a better physical understanding of the experimental data, we investigate both an adiabatic analytic model based on first-principles equations at 0 K (See supplementary section S5) in addition to macro-spin simulations that vary the temperature, nanomagnet size, anisotropy energy, and dipolar coupling strength to identify parameter space requirements for perfect signal propagation. In the macro-spin model, each magnet in the chain is represented by a single moment that possesses specific values for $K_b$ and $K_u$ that we independently vary over many simulations. Each chain is stabilized in the metastable state and simulations are run for both 0 and 300 K (at 300 K each simulation is performed 20 times). Both $K_b$ and $K_u$ are normalized by the dipolar coupling strength $M_sB$, where $M_s$ is the saturation magnetization and the dipolar coupling field $B = \frac{\mu_0}{2\pi}\frac{M_sV}{r^3}$, where $\mu_0$ is the permeability of free space, V is the nanomagnet volume, and r is the center-to-center magnet spacing. We perform simulations in a logarithmic grid of values between 0.1 and 10 for $K_b/M_sB$ and $K_u/M_sB$ and track signal propagation in each scenario. The simulations presented in Figure 4 (on a log-log scale) represent the values for $K_b/M_sB$ and $K_u/M_sB$ that produced perfect, statistically-significant, and repeatable signal propagation for 50 nm (with 20 nm spacing) and 150 nm (with 30 nm spacing) wide nanomagnets at 0 and 300 K. The color of each data point represents the average signal propagation time for each successful simulation which varies from 1.2 to 6 ns. Faster propagation times result when the anisotropy field of the easy axis increases with respect to the hard axis leading to faster magnet-to-magnet switching as expected from Gilbert damping. These plots predict a phase space of reliable propagation based on the fundamental parameters of this system.

At 0 K, the reliability phase space for both magnet sizes is similar (Figure 4a,c). The most tolerant signal propagation regions that accept large proportional deviations in both $K_b$ and $K_u$ are found in the lower left of both plots, where the coupling fields are relatively strong. This suggests that good tolerance can be experimentally realized by choosing magnetic materials with a larger $M_s$ and/or decreasing the spacing between nanomagnets. As the coupling strength in Figure 4 is reduced (upper right in each plot), we observe less tolerance for perfect signal propagation. When we incorporate stochastic temperature effects into our simulations (Figure 4b,d) we observe a reduction in the proportional tolerance for all magnet dimensions and coupling strengths. Thermal excitations increase the probability for errors to occur, diminishing the upper bound of the reliability phase space. In Figure 4b we overlay the anisotropy values and propagation distance probabilities for the ideal clocked, 300 K simulation data taken from Figures 2a and 2g which correspond to the energy range of the anisotropy engineered magnets investigated in this work. We observe that this micromagnetic range crosses a relatively narrow region of successful signal propagation that compares well with both our OOMMF simulations and experimental observations of signal propagation reliability in Figure 2. From an engineering perspective, the plots in Figure 4 provide a guide for improving the performance of future NML architectures. Fabricating notched nanomagnets with values for $K_u/M_sB$ and $K_b/M_sB$ indicated by the blue data points optimizes the speed and reliability of signal propagation. While these simulations are motivated by the lithographically tunable nanomagnets, the findings from these anisotropy models are also applicable to magnets with magnetocrystalline anisotropy[10,11]. The fixed intrinsic biaxial properties of these materials combined with nanomagnet designs less susceptible to processing variations may provide a more scalable platform to optimize signal propagation in NML chains.

In summary, this work presents the first experimental evaluation of NML operation at its ultimate intrinsic speed. By clocking NML systems at high repetition rates with nanosecond pulses, we have performed experiments that both confirm the fundamental physical mechanism of NML technology and assess NML performance with computationally-relevant cycling protocols. The dipolar coupling dynamics responsible for signal propagation in NML chains, which occurs at speeds near 100 ps per switching event, have been directly observed using time-resolved magnetic x-ray microscopy. Our computational models strongly support our experimental evidence and provide a deeper insight which will help to engineer the reliability of NML systems, where the anisotropies and dipolar coupling strengths can be optimized by a judicious choice of the nanomagnet geometry, spacing, and material.

**Methods**

*Sample preparation for MTXM*

Using electron beam lithography, e-beam evaporation, and lift-off, chains of twelve nanomagnets (2 nm Ti/12 nm permalloy, 150 nm wide, 300-360 nm long), spaced apart by 30 nm, are produced on commercially-available 500 μm thick Si wafers capped on both sides with 100 nm layers of low-stress $Si_3N_4$. Each rectangular nanomagnet has two 50 nm wide and 30 nm deep notches centrally patterned along its width. The nanomagnet chains are bound on either end by shape-biased inputs[27], which set the magnetic orientation of the first magnet in each chain, and blocks[8], which stabilize the final magnet in each chain. Next, we patterned 6 μm wide, 150 nm thick aluminum wires capped with a 10 nm layer of copper on top of the chains using optical lithography and lift-off. Finally, to create a membrane for x-ray transmission, a lithographically defined window of the backside $Si_3N_4$ layer and Si substrate is etched away releasing the remaining $Si_3N_4$ layer with the existing nanomagnet and wire features. The resistance of the Al wire after processing is 21 Ω. The membrane samples are then mounted on a Rogers 4350B printed circuit board and electrically contacted with conductive silver paint.

*MTXM experimental setup*

We used the full field transmission soft x-ray microscope (XM-1 at beamline 6.1.2) at the Advanced Light Source (ALS) at Lawrence Berkeley National Lab to obtain magnetic contrast images of our nanomagnet chains. Samples are mounted in ambient conditions at room temperature with the surface normal tilted 30° to the x-ray optical pathway to obtain a projection of the sample magnetization onto the photon propagation direction. X-ray magnetic circular dichroism at the Fe $L_3$ absorption edge (707 eV), i.e. magnetization-dependent absorption of circularly polarized x-rays transmitting through the sample, provides magnetic contrast. To enhance the magnetic contrast, each image is divided on a pixel-by-pixel basis by a reference image which is recorded in an external magnetic field strong enough to saturate the nanomagnets along their easy axes. To observe nanomagnet propagation statistics, MTXM images are repeatedly recorded after manually triggering single clocking pulses along the Al wire. Each pulse is 3 ns long with an amplitude of 18.5 V. This generates an 880 mA current which produces an on-chip magnetic field of approximately 84 mT.

*PEEM sample preparation*

Nanomagnet chains are fabricated on silicon wafers coated with a 100 nm thick insulating layer of $SiO_2$. A gold metal strip, 6 µm wide and 160 nm thick with 2 nm Ti adhesion layer is patterned on the surface using photolithography, thermal evaporation, and lift-off. A 45 nm thick layer of spin-on aluminum oxide phosphate (AlPO, Inpria Corp.) was deposited on the sample in order to smooth the surface of the Au wire on which the nanomagnet chains are fabricated. Next, chains of twelve nanomagnets (2 nm Ti/12 nm permalloy/ 2 nm Al) bounded by shape-biased inputs[27] and blocks[8] are patterned by electron beam lithography. The nanomagnets in each chain are 150 nm wide with 30 nm gaps between them, and vary in length from 300 nm to 500 nm. Each rectangular nanomagnet has two 50 nm wide and 30 nm deep notches centrally patterned along its long axis edges. A final 1 nm film of platinum is sputtered over the entire sample surface to reduce surface charging during the PEEM measurement. The resistance of the Au wire after processing is 7 Ω.

*TR-PEEM experimental setup*

Our dynamics experiment is set up as a stroboscopic pump-probe measurement where fast current pulses along the strip are synchronized with x-ray pulses from the ALS at beamline 11.0.1 during the 2-bunch operation mode. We designed and built a customized sample holder to generate the current pulses while isolating the critical electronic components from the PEEM high-voltage electron optics. The requirement in the PEEM instrument to bias the entire sample holder at high-voltage (12-18 kV) and the need for sub-nanosecond rise and fall times dictated the use of an optical link in place of electrical feedthroughs to trigger the current pulse. The in-situ pulser circuit was built from commercially-available surface mount components, an avalanche photodiode (APD), and a polyimide printed circuit board (PCB) with a silver plated copper conductor. When the APD is irradiated by a short, infrared laser pulse, it produces an electrical pulse that is amplified by two stages of radio frequency amplifiers. Current pulses with a peak amplitude of up to 1 A are delivered into the Au strip by this circuit. A near field calculation using the superposition integral formulation of the Biot-Savart law estimates that the amplified pulses produce an in-plane magnetic field with a peak amplitude of 100 mT at the location of the chains on the surface of the strip. This field is our clock mechanism (See supplementary section S4). The sample and circuit assembly contacts a copper heatsink to dissipate heat from both amplifiers. We nominally operate this measurement at room temperature inside the PEEM vacuum chamber. The probe beam consists of 70 ps x-ray pulses at a repetition rate of 3 MHz. 850 nm pulses with 1-2 ns duration from a fast diode laser strike the APD at the same repetition rate. The diode laser is mounted outside the PEEM vacuum chamber and the beam is free-space coupled through a window port and focused onto the APD. A pulse delay generator triggered by a synch pulse from the synchrotron RF system synchronizes the x-ray bunches to the laser pulse. Accumulating magnetic contrast images in the PEEM for varying delay allows us to study the time dynamics.

As with the MTXM images, we obtain magnetic contrast by XMCD tuning the x-rays to the Fe $L_3$ absorption edge, expose two images at the same location illuminated by right and left circularly polarized x-rays, respectively, and then compare them by per-pixel numerical division. The images are then adjusted using a median noise filter and linear brightness and contrast stretching calibrated with nearby non-magnetic regions. We imaged 22 chains and each image is averaged over 360 million clock cycles (2 minutes integration time).

**Acknowledgements**

We gratefully acknowledge support from the Western Institute of Nanoelectronics, DARPA, and the NSF Center for Energy Efficient Electronics Science. Work at the Advanced Light Source, Center for X-ray Optics, and the Molecular Foundry at Lawrence Berkeley National Laboratory is supported by the Director, Office of Science, Office of Basic Energy Sciences, US Department of Energy under contract number DE-AC02-05CH11231. M.-Y. I. and P.F. acknowledge support by the Leading Foreign Research Institute Recruitment Program (Grant No. 2012K1A4A3053565) through the National Research Foundation of Korea (NRF) funded by the Ministry of Education, Science, and Technology (MEST). In addition, we acknowledge the Marvell Nanofabrication Laboratory for the cleanroom and machine shop facilities.

**Author contributions**

MTXM samples were prepared by Z.G., D.B.C., and W.C. MTXM measurements were performed by Z.G., M.-Y. I., and P.F. TR-PEEM samples were prepared by Z.G., D.B.C., W.C., and P.B. TR-PEEM experiments were performed by Z.G., M.E.N., R.S., J.H., B.L., M.T.A., M.A.M., A.D., A.Y., and A.S. Simulations were performed by Z.G. and B.L. The experiments were conceived by J.B. and the manuscript was written by M.E.N., Z.G., and J.B. All authors discussed the results and commented on the manuscript.

**Competing financial interests**

The authors declare no competing financial interests.

**Figures**

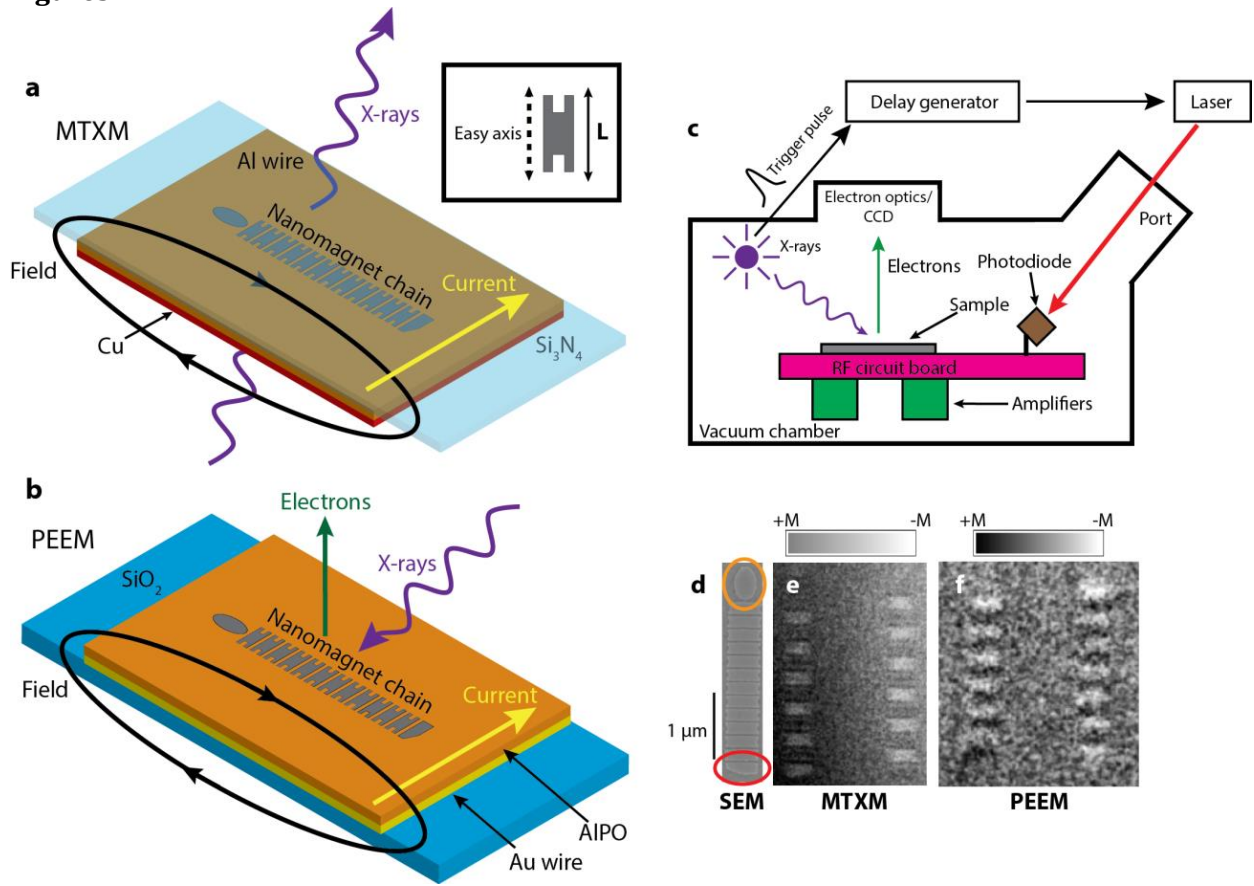

**Figure 1: Sample design and x-ray microscopy experimental setup.** a,b) Schematic of soft x-ray transmission and photo-emission electron microscopy experiments. In both experiments nanomagnet chains are fabricated on metal wires. Current pulses generating on-chip fields reset the orientation of all nanomagnets after each clocking cycle. Inset: orientation of the easy axis for a nanomagnet of length, L, with biaxial stability. c) Schematic of TR-PEEM setup. A trigger pulse generated by the ALS excites laser pulses which are focused on the photodiode of a customized circuit board containing the sample inside the PEEM vacuum chamber. The circuit amplifies the current pulses from the photodiode to generate on-chip clocking fields which are synchronized with x-ray pulses by a delay generator. d) SEM of chain with 12 nanomagnets. The input and terminating block magnets are circled in red and orange, respectively. e,f) Magnetic contrast images of chains with 12 nanomagnets observed by MTXM and PEEM.

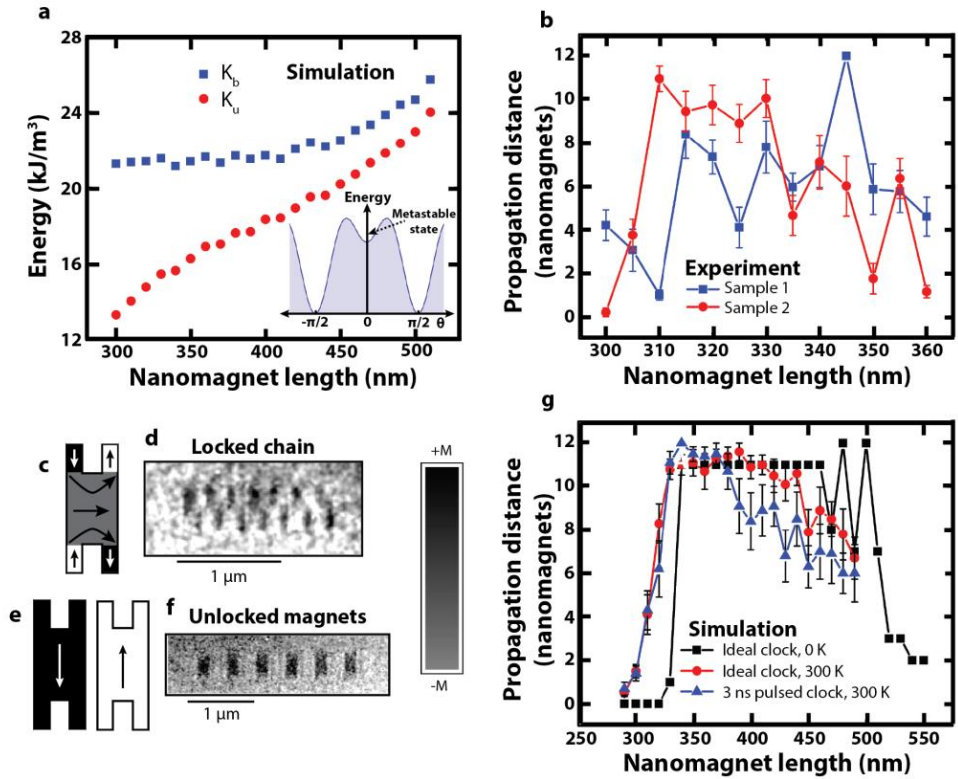

**Figure 2: Identifying regions of optimal NML signal propagation for pulsed clocking fields.** a) The uniaxial ($K_u$) and biaxially ($K_b$) anisotropy energy as a function of nanomagnet length for 150 nm wide nanomagnets calculated from an analytical model and OOMMF simulations. Inset: schematic energy diagram with respect to the magnetic orientation from the hard axis indicating the biaxial anisotropy-generated metastable state. b) Signal propagation distance in nanomagnet chains from the input magnet as a function of nanomagnet length measured in two samples with MTXM. Error bars report the standard error of the mean. c,d) Illustration and MTXM image of a locked chain of 7 nanomagnets. The spatial resolution of MTXM is sufficient to distinguish the domains in a biaxially engineered nanomagnet. e,f) Illustration and MTXM image of perfect signal propagation in a chain of 12 nanomagnets. g) Signal propagation distance of nanomagnet chains from the input magnet as a function of nanomagnet length calculated from OOMMF simulations for ideally initialized chains clocked at 0 K (black squares), 300 K (red circles), and chains initialized by a 3 ns clocking pulse at 300 K (blue triangles). Error bars report the standard error of the mean.

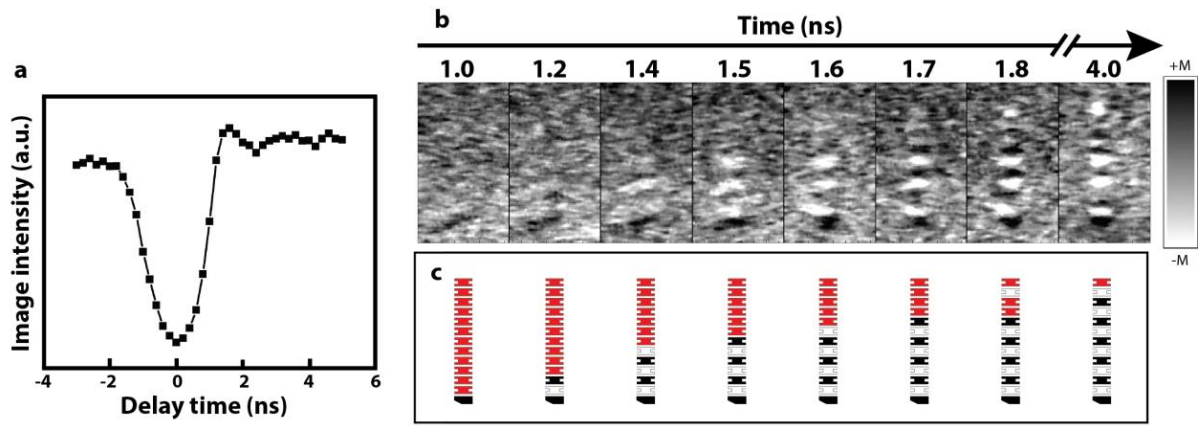

**Figure 3: Observation of signal propagation in an NML chain by dipolar coupling.** a) Photo-electron intensity count for an image containing the Au wire vs. time delay. b) Averaged time-resolved XMCD-PEEM images of a nanomagnet chain at various time delays from 1 to 4 ns. The chain demonstrates behavior which suggests dipolar coupling is switching magnets sequentially at fast timescales on the order of 100 ps. c) The authors' interpretation of the switching events observed in part b). Red indicates magnets in their metastable state after clocking and black and white indicate magnets that have oriented along their easy axis in opposite directions, respectively.

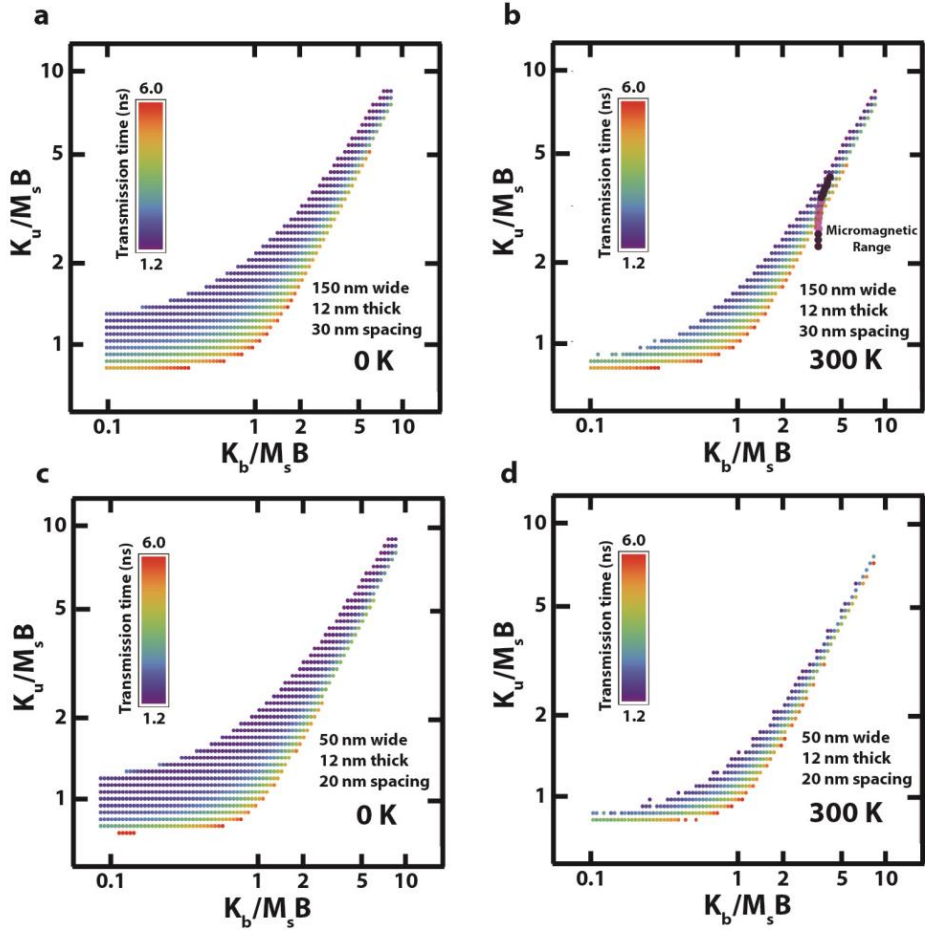

**Figure 4: Reliability phase space calculated by macro-spin simulations and comparison with micromagnetic simulations.** Log-log macro-spin simulation plots for 12 magnet chains with 150 nm wide nanomagnets at a) 0 K and b) 300 K and 50 nm wide nanomagnets at c) 0 K and d) 300 K that indicate values for $K_u$ and $K_b$ (normalized to the saturation magnetization, $M_s$, and dipolar field coupling, *B*) that produce repeatable and reliable signal propagation starting from an ideally initialized metastable state. Simulations run at 300 K are performed 20 times per data point to statistically account for stochastic thermal effects. The color scale indicates the average signal propagation time to complete each successful simulation. The anisotropy values from Figure 2a and micromagnetic simulation signal propagation distance from Figure 2g for ideal clocking at 300 K are overlaid onto part b). The magenta coloring in these points indicates higher signal propagation distance.

Supplementary information to accompany:

# Sub-nanosecond signal propagation in anisotropy engineered nanomagnetic logic chains

Zheng Gu[1,†], Mark E. Nowakowski[1,†], David B. Carlton[2], Ralph Storz[3], Mi-Young Im[4,5], Jeongmin Hong[1], Weilun Chao[4], Brian Lambson[6], Patrick Bennett[1], Mohmmad T. Alam[7], Matthew A. Marcus[8], Andrew Doran[8], Anthony Young[8], Andreas Scholl[8], Peter Fischer[4,9], and Jeffrey Bokor[1,*]

[1]*Department of Electrical Engineering and Computer Sciences, University of California, Berkeley, California 94720, USA*
[2]*Intel Corp., 2200 Mission College Blvd., Santa Clara, California 95054, USA*
[3]*Thorlabs Inc., 56 Sparta Ave., Newton, New Jersey 07860, USA*
[4]*Center for X-ray Optics, Lawrence Berkeley National Laboratory, Berkeley, California 94720, USA*
[5]*Daegu Gyeongbuk Institute of Science and Technology, Daegu 711-873, Korea*
[6]*iRunway, 2906 Stender Way, Santa Clara, California 95054, USA*
[7]*Intel Corp., 5200 NE Elam Young Pkwy, Hillsboro, OR 97124, USA*
[8]*Advanced Light Source, Lawrence Berkeley National Laboratory, Berkeley, California 94720, USA*
[9]*Department of Physics, University of California, Santa Cruz, California 94056, USA*

† Denotes equal contributions
* Corresponding author

**Contents**

**Section S1: Design of input and block magnets**

**Section S2: Calculating $K_b$ and $K_u$ using an analytical model and OOMMF**

**Section S3: OOMMF simulation of signal propagation in longer chains**

**Section S4: TR-PEEM magnetic contrast images indicate successful pulse clocking**

**Section S5: Calculating clocking stability and transmission reliability analytically**

**Section S1: Design of input and block magnets**

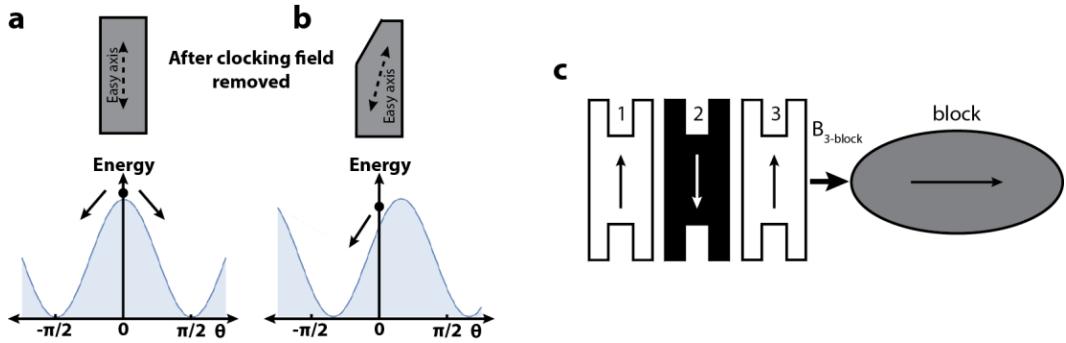

Figure S1: a,b) Schematic and energy diagrams of an input magnet without (a) and with shape anisotropy. The black dot indicates the magnetization orientation immediately after the clocking field is removed. c) A chain of nanomagnets are terminated by a 'block' nanomagnet.

*Input magnets*

Ensuring repeatable performance in our magnet chains for the stroboscopic time-resolved measurement was critical. We required the behavior of each individual magnet to be identical each clock cycle. To accomplish this, the easy axis of each input magnet was rotated with shape anisotropy engineering such that a component of the easy axis was projected along the clocking field direction. After the clocking field is removed an input magnet with shape anisotropy (Figure S1b) strongly prefers to relax into only one easy axis state; for contrast an input magnet without shape anisotropy (Figure S1a) would have an equal probability of relaxing into either degenerate easy axis state. [Main article ref. 27]

*Terminating 'block' magnets*

The final magnet in the chain requires a special terminating nanomagnet neighbor called a block. In the absence of this block a final magnet with only one neighbor is susceptible to relaxing out-of-sequence due to an absence of symmetric hard axis coupling fields which typically provide stability in magnets with two neighbors (See supplementary section S5). To add stability to the final magnet, an ellipse-shaped block magnet with an easy axis parallel to the clocking field direction is fabricated next to the final magnet to provide an additional coupling field $B_{3\text{-block}}$ as shown in Figure S1c. [Main article ref. 8]

## Section S2: Calculating $K_b$ and $K_u$ using an analytical model and OOMMF

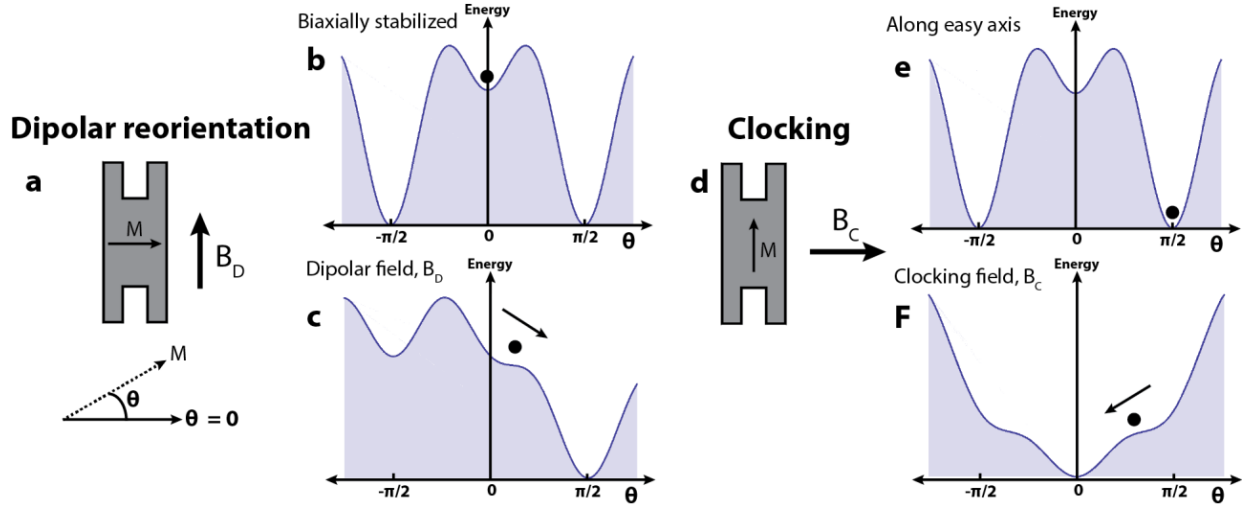

Figure S2: a) Magnetic orientation for driving moments stabilized along the hard axis to the easy axis with a dipolar coupling field ($B_D$). b) Energy diagram of the starting orientation for a biaxially stabilized magnet. c) The saddle point defined at $B_D$ when the magnet reorients from the hard axis to the easy axis. d) Magnetic orientation for driving moments stabilized along the easy axis to the hard axis with an external clocking field ($B_C$). e) Energy diagram of the starting orientation for a magnet oriented along its easy axis. f) The saddle point defined at $B_C$ when the magnet reorients from the easy axis to the hard axis.

The values of the biaxial ($K_b$) and uniaxial ($K_u$) anisotropy energy in Figure 2a are estimated using both an analytical model and micromagnetic simulations. Using an analytical model, we first calculate the relationship between the magnetic dipolar coupling field ($B_D$) and the clocking field ($B_C$) as a function of $K_b$, $K_u$, and the magnetization saturation ($M_S$). To determine these functions we consider two cases: i) a nanomagnet is stabilized in the biaxial metastable state and a dipolar coupling field ($B_1$) from a neighboring magnet is applied perpendicular to the magnetization (Figure S2a) and ii) a nanomagnet is stabilized along its easy axis and an external clocking field ($B_2$) is applied perpendicular to the magnetization (Figure S2d).

The energy equation describing the first scenario is

$$U = -K_u \sin^2 \theta - \frac{K_b}{4} \cos^2 2\theta - M_S B \sin \theta \quad (1).$$

Assuming $K_b > K_u$ the moment can be stabilized in a metastable region along the hard axis as shown in Figure S2b, where $\theta = 0$ corresponds to the hard axis. Applying a dipolar field $B_1$ tilts the energy landscape until the energy barrier defined by $K_b$ and $K_u$ becomes a saddle point at $\frac{dU}{d\theta} = 0$ and $\frac{d^2U}{d\theta^2} = 0$ (Figure S2c). Solving these equations for $B_1$ in the first quadrant gives

$$B_D = \frac{-4}{3M_S} \sqrt{\frac{(K_b - K_u)^3}{6K_b}} \quad (2),$$

where $B_D$ is the value of $B_1$ at the saddle point. In this case $K_b \geq K_u$ is required to obtain a real solution, otherwise no saddle point emerges since there is no energy barrier.

The energy equation describing the second scenario is

$$U = -K_u \cos^2 \theta - \frac{K_b}{4} \cos^2 2\theta - M_S B \sin \theta \quad (3).$$

The moment starts stabilized along the easy axis as shown in Figure S2e, where $\theta = 0$ corresponds to the hard axis. Applying a clocking field $B_2$ tilts the energy landscape until the energy barrier defined by $K_b$ and $K_u$ becomes a saddle point at $\frac{dU}{d\theta} = 0$ and $\frac{d^2U}{d\theta^2} = 0$ (Figure S2f). Solving these equations for $B_2$ in the first quadrant gives

$$B_C = \frac{4}{3M_S} \sqrt{\frac{(K_b + K_u)^3}{6K_b}} \quad (4),$$

where $B_C$ is the value of $B_2$ at the saddle point. In this case there is no requirement of $K_b \geq K_u$ because there is always an energy barrier going from the easy axis to the hard axis due to $K_u$.

We calculate $B_D$ and $B_C$ for the nanomagnet geometries used in our experiments using micromagnetic simulations in OOMMF. Within the geometries simulated, the largest $B_C$ was 45 mT, meaning that clocking fields exceeding 45 mT by a reasonable margin are sufficient. $K_b$ and $K_u$ for each geometry are back calculated using equations 2 and 4. This assumes the energy barrier defined by $K_b$ and $K_u$ can be traversed from both the easy axis and metastable hard axis. This biaxial anisotropy approximation addresses the essential elements of NML: determining the field energy required via dipolar coupling and external clocking to reorient moments between easy and hard axes over many cycles.

## Section S3: OOMMF simulation of signal propagation in longer chains

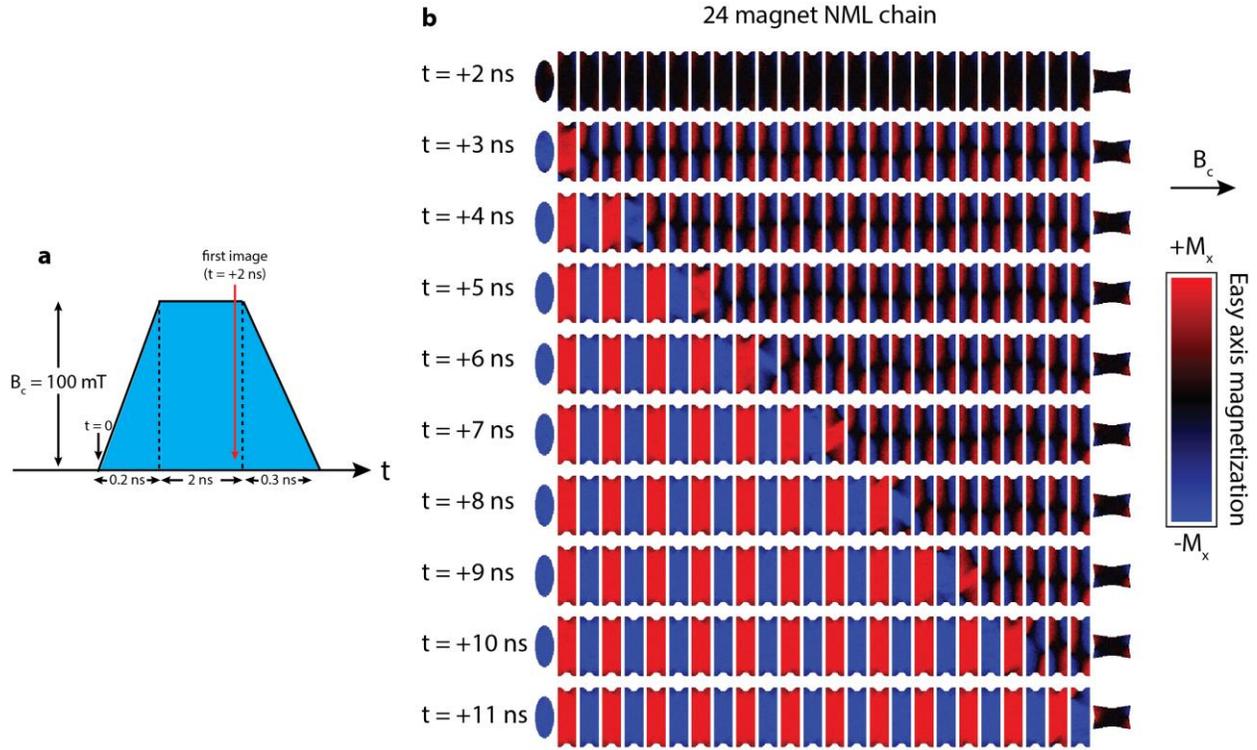

Figure S3: a) Clocking pulse profile applied prior to behavior observed in part b). b) Perfect signal propagation in a chain with 24 nanomagnets with engineered biaxial anisotropy, simulated at room temperature.

To demonstrate the superior stability gained by engineering a metastable state with biaxial anisotropy, we simulate signal propagation in a chain twice as long as the chains we measured with TR-PEEM. Each magnet (excluding the input and the block) in the 24 magnet chain of Figure S3b is configured with dimensions identical to those used in the TR-PEEM experiment: 450 nm x 150 nm and 12 nm thick. The OOMMF simulation (at T = 300 K) is initialized with a trapezoidal-shaped 2 ns clocking field pulse ($B_c$) of 100 mT (Figure S3a). This is similar to the one used in the simulations of Figure 2g of the main article. After the clocking pulse is removed we confirm each individual nanomagnet remains stability oriented along its hard axis until it is excited by its left-most nearest neighbor. This simulation, which depicts perfect signal propagation in ambient conditions, demonstrates that with a judicious choice of biaxial anisotropy accurate propagation is achievable in chains of arbitrary lengths.

**Section S4: TR-PEEM magnetic contrast images indicate successful pulse clocking**

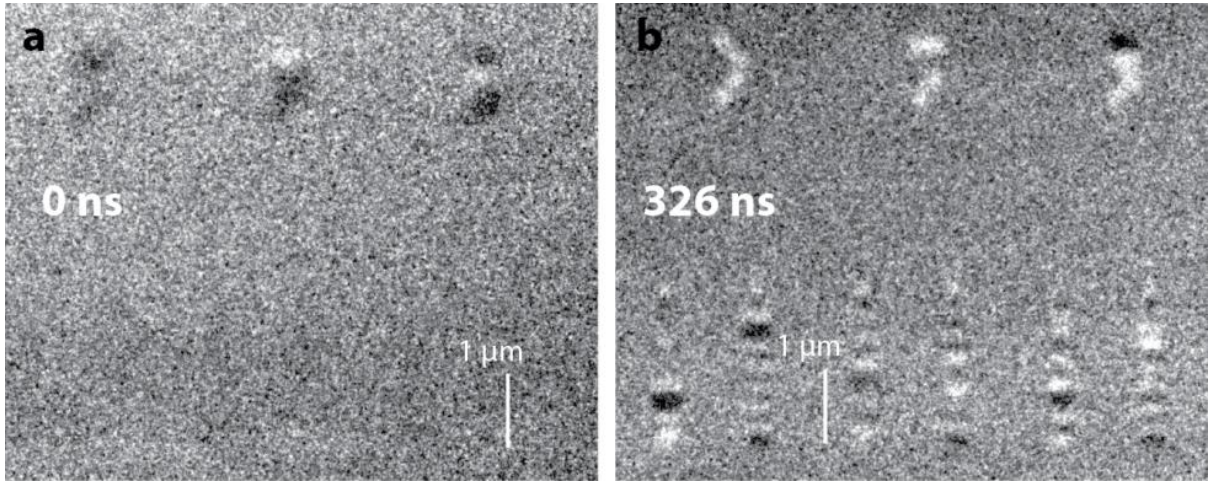

Figure S4. a) XMCD-PEEM image taken at 0 ns delay time. Index magnets show magnetic contrast. Magnets along the wire are oriented along their hard axis and are perpendicular to the magnetic contrast direction. b) XMCD-PEEM image taken at 326 ns delay time. Index magnets show magnetic contrast. Magnets along the wire are oriented along their easy axis and are parallel to the magnetic contrast direction.

During our time-resolved PEEM measurement we vary the delay time between the clocking pulse and the x-ray pulse with a delay generator. Figure 3a in the main manuscript characterizes the clocking pulse by measuring the photo-electron intensity of a PEEM image as a function of delay time. During the rising and falling edges images appear blurry and move due to the Lorentz force acting on the electrons and the time averaging of jitter. However at the peak of the pulse, images are stable and we observe magnetic contrast of index magnets fabricated off of the wire designed to indicate the specific nanomagnet length in each chain (Figure S4a). Along the wire we observe no magnetic contrast which indicates that during the pulse all nanomagnets are aligned along their hard axis, perpendicular to the XMCD contrast direction. Here, we compare an image taken at the pulse peak with an image taken at 326 ns after the pulse peak (Figure S4b). We also observe that the pulse field can influence the orientation of the indexing magnets off the wire.

**Section S5: Calculating clocking stability and transmission reliability analytically**

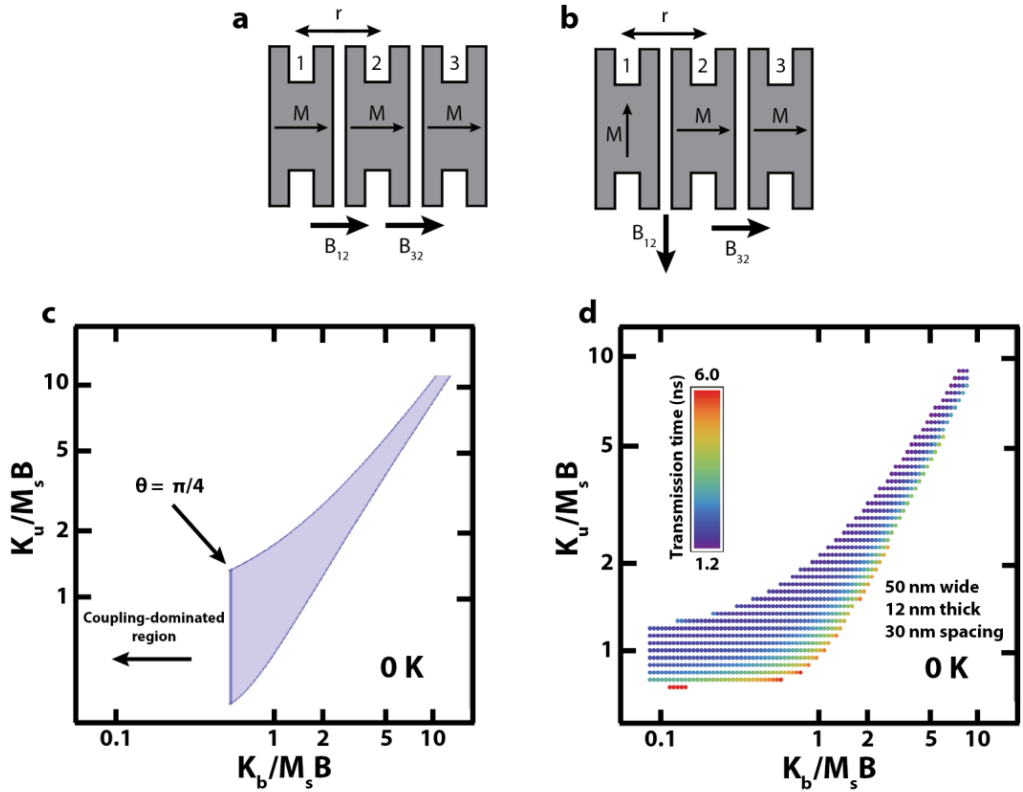

Figure S5: a) Magnetic orientation for 3 nanomagnets stabilized along their hard axis. Magnet 1 and 3 act on magnet 2 via their dipolar fields ($B_{12}$ and $B_{32}$) which are parallel to the magnetization in the metastable state. b) Magnetic orientation for 3 nanomagnets during signal transmission, where signal is propagating from magnet 1 to 3. The dipolar field $B_{12}$ is now oriented perpendicular to the metastable magnetization state. c) Analytical calculation, performed at 0 K, of a reliable transmission region based on the geometry of part b) plotted on a log-log scale. The coupling-dominate region is indicated, but not plotted (see text). d) Macro-spin simulation plot for a 12 magnet chain with 50 nm wide nanomagnets at 0 K plotted on a log-log scale. A coupling-dominate region is predicted and illustrated.

*Clocking stability*

To analytically derive a condition for clocking stability we consider three closely spaced magnets (Figure S5a) with both uniaxial ($K_u$) and biaxial anisotropy ($K_b$) separated by a center-to-center distance, *r*, and calculate the influence of dipolar fields ($B_{12}$ and $B_{32}$) from the edge magnets (1 and 3) on the central magnet (2). Assuming the magnets are initialized by an external clocking field oriented along their hard axes, we calculate the maximum value of $K_u$ for the three magnets, such that magnet 2 remains in the metastable state.

The total energy of magnet 2, including both uniaxial and biaxial anisotropy terms and the dipolar fields from both neighboring magnets is

$$U = -K_u \sin^2\theta - \frac{K_b}{4}\cos^2 2\theta - 2M_S B \cos\theta \quad (5),$$

where $B$ is the dipolar coupling field from one neighboring magnet ($B = B_{12} = B_{32}$),

$$B = \frac{\mu_0}{4\pi}\left(\frac{3\vec{r}\cdot(\vec{m}\cdot\vec{r})}{r^5} - \frac{\vec{m}}{r^3}\right) = \frac{\mu_0 M_S V}{2\pi r^3} \quad (6),$$

given $\vec{r}\cdot(\vec{m}\cdot\vec{r}) = mr^2$ for a nanomagnet volume $V$. For magnet 2 to remain magnetized along its hard axis there must be an energy barrier defined by $K_b$ and $K_u$. If $K_u$ is raised with respect to $K_b$ the energy barrier is reduced. At $\frac{dU}{d\theta} = 0$ and $\frac{d^2U}{d\theta^2} = 0$ the energy barrier becomes a saddle point which allows spontaneous realignment from the hard axis to the easy axis. Solving these equations gives

$$K_u < K_b + M_S B \quad (7)$$

which defines conditional bounds for $K_u$ and $K_b$ with respect to $B$ for stable clocking into the metastable state.

*Transmission reliability*

To analytically derive a condition for transmission reliability we consider three closely spaced magnets (Figure S5b) with identical parameters as above, however now we assume transmission propagation is occurring (after stable clocking) from magnet 1, which has been reoriented along its easy axis. Again, we calculate the influence of dipolar fields ($B_{12}$ and $B_{32}$) from the edge magnets (1 and 3) on the central magnet (2) to calculate the parameter requirements necessary for dipolar field realignment in magnet 2.

Because $B_{12}$ and $B_{32}$ are no longer equal:

$$B_{12} = \frac{\mu_0}{4\pi}\left(\frac{3\vec{r}\cdot(\vec{m}\cdot\vec{r})}{r^5} - \frac{\vec{m}}{r^3}\right) = \frac{\mu_0 M_S V}{2\pi r^3} \quad (8), \text{ given } \vec{r}\cdot(\vec{m}\cdot\vec{r}) = mr^2 \text{ and}$$

$$B_{32} = \frac{\mu_0}{4\pi}\left(\frac{3\vec{r}\cdot(\vec{m}\cdot\vec{r})}{r^5} - \frac{\vec{m}}{r^3}\right) = -\frac{\mu_0 M_S V}{4\pi r^3} \quad (9), \text{ given } \vec{m}\cdot\vec{r} = 0,$$

the total energy equation is now given by

$$U = -K_u \sin^2 \theta - \frac{K_b}{4}\cos^2 2\theta - M_S B \left(\cos \theta + \frac{\sin \theta}{2}\right) \qquad (10).$$

For magnet 2 to reorient, the energy barrier must be removed by the dipole fields from its neighbors. This occurs at a saddle point in the energy equation when $\frac{dU}{d\theta} = 0$ and $\frac{d^2U}{d\theta^2} = 0$ between $0 < \theta \leq 0.228\pi$. Solving these equations gives two expressions which define a region for transmission reliability in this system:

$$\frac{8K_b}{M_S B} < \frac{1}{2\sin^3 \theta} + \frac{1}{\cos^3 \theta} \quad (11) \text{ and}$$

$$\frac{8K_u}{M_S B} > \frac{1}{2\sin^3 \theta} - \frac{1}{\cos^3 \theta} - \frac{3}{\sin \theta} + \frac{6}{\cos \theta} \quad (12).$$

Within this region, transmission along a chain behaves as expected: magnet 2 will remain in the metastable state until it is driven by the dipolar fields of magnet 1 to its easy axis. We note that this model is adiabatic and is only valid at 0 K.

We also note that our calculation limitation of $0 < \theta \leq 0.228\pi$ suggests that there is no threshold past $\theta = \frac{\pi}{4}$. This is because $\frac{1}{2\sin^3 \theta} + \frac{1}{\cos^3 \theta}$ reaches its minimum value at $\theta \approx 0.228\pi$, however it is possible to choose smaller values for $K_b$, but there will be no saddle point, despite this magnetic reorientation is still permitted because a local energy minimum is present. Due to the small values of $K_b$ relative to $M_s B$, we call this region coupling-dominated, where biaxial anisotropy is weak compared to dipolar coupling and no energy barrier impedes switching. Plotting equations 7, 11, and 12 illustrates a region of permissible parameters ($K_b$, $K_u$, $M_s$, and B) for reliable transmission (Figure S5c). We compare this analytical plot to a plot generated using macro-spin (Figure S5d) which calculated stable transmission in chains of 12 closely spaced nanomagnets as described in the main paper. The coupling-dominated region is apparent in the macro-spin simulation.

The qualitative and quantitative nature of Figure S5c and S5d improve the confidence of our macro-spin simulations. Our analytical model is two-dimensional, adiabatic, and is performed at 0 K. The simulations assume three-dimensional properties, time-dependent dynamics based on the Landau-Lifshitz-Gilbert (LLG) equation, and are performed at 0 and 300 K. Comparing these two simulations suggests a physically consistent understanding (even with a single macro-spin approximation) of the fundamental properties and performance of this system. Based on this understanding we are able to learn more about potential error mechanisms (e.g. non-nearest neighbor dipolar coupling which cause magnets two or more positions distant to reorient, or the effects of thermal fluctuations) and can make better design choices in future attempts to demonstrate improved reliability.